\documentclass[12pt]{iopart}

\usepackage{amsfonts}
\newcommand{\beq}{\begin{equation}}
\newcommand{\eeq}{\end{equation}}
\newcommand{\la}{\langle}
\newcommand{\ra}{\rangle}
\newcommand{\x}{\overline{x}}
\newcommand{\p}{\overline{p}}

\begin{document}
\title[Quantum Langevin equation]
{Quantum Langevin equation}

\author{M\'{a}rio J. de Oliveira}

\address{Instituto de F\'{\i}sica,
Universidade de S\~{a}o Paulo, \\
Rua do Mat\~ao, 1371,
05508-090 S\~{a}o Paulo, S\~{a}o Paulo, Brazil}

\ead{oliveira@if.usp.br}

\begin{abstract}

We propose a Langevin equation to describe the quantum Brownian 
motion of bounded particles based on a distinctive formulation
concerning both the fluctuation and dissipation forces.
The fluctuation force is similar to that employed in the
classical case. It is a white noise with a variance proportional
to the temperature. The dissipation force is not restrict to
be proportional to the velocity and is determined in a way as to guarantee that
the stationary state is given by a density operator of the Gibbs
canonical type. To this end we derived an equation that gives the
time evolution of the density operator, which turns out to be a
quantum Fokker-Planck-Kramers equation. The approach is applied to the
harmonic oscillator in which case the dissipation force is found to be
non Hermitian and proportional to the velocity and position. 

\end{abstract}

\maketitle

%--------------------------------------------------------------------
\section{Introduction}

The Brownian motion of particles in contact
with a heat reservoir is well described by the Langevin
equation \cite{kampen1981,risken1989,gardiner2009,coffey1996}.
The essential feature of this equation is the presence of
two types of forces, one being the dissipation force and the other being
a force of a stochastic nature. The interplay between these
two forces leads to a mechanism of loss and gain of energy
that ensures an unceasing motion of the particles. In addition,
a dissipation-fluctuation relation exists or should 
exist between the two forces to guarantee that in equilibrium
the probability distribution of velocities and positions
is the equilibrium Gibbs distribution.

The extension of the Langevin equation to describe
quantum open system and quantum dissipation 
\cite{carmichael1999,breuer2002,gardiner2010,weiss2012,caldeira2014}
has been addressed but the quantum description has brought
problems that were not present in the classical case,
and the approach to quantum open system and quantum dissipation
does not seem to be completely solved \cite{araujo2019}.
For instance, an equation that governs the time evolution
of the probability density, the Fokker-Planck-Kramers equation, can 
immediately be obtained from the classical Langevin equation.
However, for the quantum case, such a derivation is not a trivial
problem \cite{araujo2019}.

The development of an equation that describes quantum dissipation
has focused mainly on the harmonic oscillator.  This is the case
of the paper by Senitzky  on quantum dissipation
\cite{senitzky1960,senitzky1961,frenkel2012}. He remarked that in quantum
mechanics, dissipation and fluctuations cannot be divorced as in
classical mechanics so that these two mechanisms must be considered
together. In his treatment of the harmonic oscillator the dissipation
force is assumed to be proportional to the velocity, and the 
fluctuating force has a time correlation proportional to 
$E(\omega)\delta(t'-t)$ where $\omega$ is the oscillator frequency and 
\beq
E = \frac{\hbar\omega}2 \coth\frac{\hbar\omega}{2k_BT}
\eeq
is the Planck function, the average energy of a harmonic oscillator.
Another type of time correlation for the fluctuating force
was that used by Ford, Kac and Mazur
\cite{ford1965,pasquale1984,ford1987,ford1988}.
They proposed a symmetrized time correlation proportional to 
\beq
\int_0^\infty E(\omega) \cos\omega(t'-t) d\omega.
\eeq
This type of correlation has been used to study the thermal properties
of chains coupled oscillator in equilibrium and out of equilibrium
\cite{zurcher1990}.
However, it has been shown that the canonical distribution cannot be approached
for finite friction but only in the weak friction limit
\cite{benguria1981,maassen1984}.

The approach proposed by Bedeaux and Mazur \cite{bedeaux2001} 
for the Langevin equation for a harmonic oscillator has
a distinct feature concerning the fluctuation mechanism.
In addition to the random force, they introduced a random velocity.
The correlation between the random force and the random velocity yields the
zero point motion of the oscillator. The correlation of the random
force is similar to that employed by Senitzky, $E(\omega)\delta(t'-t)$,
and the correlation of the random velocity is chosen to vanish. However,
the correlation between the random force and random velocity
is nonzero and is proportional to $i\hbar\delta(t'-t)$. This
approach has recently been used to an axiomatic construction of
quantum Langevin equation \cite{araujo2019}.

Here we propose an approach to the Langevin equation that is distinct
from the above approaches in two aspects. In treatments of
the Langevin equation, it is usual to consider the
dissipative force as proportional to the velocity.
In our approach, this restriction is relaxed and we
allow for a dissipative force that might depend on velocity and 
possibly on the position. The second distinct feature concerns the
random force. The correlation of the random force is taken to
be proportional to the temperature, as in the classical case,
and is considered to be a complex scalar. These two features
allowed us to derive the following relation between the random
dissipative force $F$ and the momentum $p$ and the position $x$ of a particle
described by a Hamiltonian ${\cal H}(x,p)$,
\beq
F = \gamma \{ p - \frac{\beta}{2!}[p,{\cal H}]
+ \frac{\beta^2}{3!}[[p,{\cal H}],{\cal H}] + \ldots \},
\label{7}
\eeq
where $\gamma$ is the dissipative constant and $\beta=1/k_BT$. 
This expression is non Hermitian in general and
has been previously obtained by another procedure
namely by the quantization of the classical Fokker-Planck-Kramers
equation \cite{oliveira2016}. Expression (\ref{7}) gives a dissipation
proportional to the form $F=\gamma p$ only for a free
Brownian motion in which case ${\cal H}=p^2/2m$. In the
classical limit, it reduces to the usual form $F=\gamma p$.

The present approach allows us to derive the
equation that gives the evolution of the density operator,
which is a quantum Fokker-Planck-Kramers equation \cite{oliveira2016}.
The stationary solution of this equation shows that the
system reaches indeed the appropriate Gibbs state, that is, the system
thermalizes properly. We remark that this is valid for any finite 
dissipative constant and not only for the weak friction limit
and for any potential and not for the harmonic potential as long as
the expression (\ref{7}) is used.

Our approach follows as much as possible the usual treatments of
the classical Langevin equation. For this reason we will first
review briefly the classical approach to the Langevin equation.
After that we present our approach to the quantum stochastic equation
and derive from it the quantum Fokker-Planck-Kramers equation.
In the last section before the conclusion we apply our
approach to the harmonic oscillator
in which case the expression (\ref{7}) 
reduces to the form $F=\gamma(ap+ibx)$.

%--------------------------------------------------------------------
\section{Classical stochastic equation}

We consider the Browian motion of a bounded particle of mass $m$
moving along a straight line and denote by $x$ its position and
by $p$ is momentum. The Langevin equation for this particle is
\beq
\frac{dp}{dt} = - \frac{\partial{\cal H}}{\partial x} - F + {\mathfrak F},
\label{21b}
\eeq
\beq
\frac{dx}{dt} = \frac{\partial{\cal H}}{\partial p},
\label{21a}
\eeq
where ${\cal H}(x,p)$ is the Hamiltonian function
\beq
{\cal H} = \frac{p^2}{2m}  + V(x),
\eeq
$F$ is a dissipative force and ${\mathfrak F}$ is
a fluctuating force holding the properties $\la{\mathfrak F}(t)\ra=0$ and
\beq
\la{\mathfrak F}(t){\mathfrak F}(t')\ra = \Gamma\delta(t-t').
\eeq
We consider $\Gamma$ to be constant, independent of $x$ and $p$. 

In the stationary regime we require that
the particle be in thermodynamic equilibrium with its surroundings,
which is assumed to be at at temperature $T$. The requirement is
fullfiled by demanding that, in the stationary regime,
the probability distribution of the variables $x$ and $p$
is described by the Gibbs canonical distribution
\beq
\rho_0 = \frac1\zeta e^{-\beta{\cal H}},
\label{24}
\eeq
where $\beta=1/k_BT$, a condition implying that dissipation and fluctuation are
related in some way. One procedure used to find such a relationship
amounts to write down the time evolution equation for the probability 
distribution and impose that, in the stationary regime it is of the form (\ref{24}).

To find the equation that gives the time evolution of the probability
distribution, we start by writing
the Langevin equations (\ref{21b}) and (\ref{21a}) in the discretized form 
\beq
\Delta p
= - \frac{\partial{\cal H}}{\partial x}\Delta t - F \Delta t + \xi \sqrt{\Gamma \Delta t},
\label{19b}
\eeq
\beq
\Delta x = \frac{\partial{\cal H}}{\partial p}\Delta t,
\label{19a}
\eeq
where $\xi$ is a random variable with central Gaussian distribution
with unit variance.

The variation of an arbitrary function ${\cal F}(x,p)$ is given by
\beq
\Delta {\cal F} = \frac{\partial {\cal F}}{\partial x} \Delta x 
+ \frac{\partial {\cal F}}{\partial p} \Delta p
+ \xi^2\frac\Gamma{2} \frac{\partial^2{\cal F}}{\partial p^2} \Delta t.
\eeq
Replacing $\Delta x$ and $\Delta p$, given by the equations (\ref{19a})
and (\ref{19b}) into this equation, 
and after integrating in $\xi$ and dividing both sides by $\Delta t$,
the following equation is obtained by taking the limit $\Delta t\to0$,
\beq
\frac{\partial{\cal F}}{\partial t}
= \{{\cal F},{\cal H}\}-\frac{\partial{\cal F}}{\partial p} F
+ \frac{\Gamma}2\frac{\partial^2{\cal F}}{\partial p^2},
\eeq
where
\beq
\{{\cal F},{\cal H}\}
= \frac{\partial{\cal F}}{\partial x}\frac{\partial{\cal H}}{\partial p}
- \frac{\partial{\cal H}}{\partial x}\frac{\partial{\cal F}}{\partial p}.
\eeq
Denoting by $\rho(x,p)$ the probability distribution of variables $x$ and $p$,
the time evolution of the average $\la{\cal F}\ra$ becomes
\beq
\frac{d}{dt} \la{\cal F}\ra
= \int\left(\{{\cal F},{\cal H}\} - \frac{\partial{\cal F}}{\partial p} F
+ \frac{\Gamma}2 \frac{\partial^2{\cal F}}{\partial p^2}\right)\rho\, dxdp.
\eeq
Since ${\cal F}$ is arbitrary, we obtain from this equation the following 
evolution of the probability distribution, after appropriate integrations
by parts, and considering that the density vanishes at the boundaries
of the system,
\beq
\frac{\partial\rho}{\partial t}
= \{{\cal H},\rho\} + \frac{\partial F \rho}{\partial p} 
+ \frac\Gamma2 \frac{\partial^2\rho}{\partial p^2},
\eeq
which is the Fokker-Planck-Kramers equation. 

Imposing that the stationary probability distribution is the Gibbs distribution
(\ref{24}), and using the result $\{{\cal H},\rho_0\}=0$, we reach the condition 
\beq
F\rho_0 + \frac \Gamma2\frac{\partial\rho_0}{\partial p} = 0.
\label{12}
\eeq
Replacing the Gibbs distribution (\ref{24}) into 
this equation to find
\beq
F = \frac{\Gamma\beta}{2m} p,
\label{20}
\eeq
which is the desired relation between  dissipation and fluctuation.
If we write $F=\gamma p$, where $\gamma$ is the dissipation constant,
we reach the usual dissipation-fluctuation relation
\beq
\Gamma = \frac{2m \gamma}{\beta}.
\eeq

%--------------------------------------------------------------------
\section{Quantum stochastic equation}

In analogy with the classical equations of motion (\ref{21b}) and (\ref{21a}),
we assume the same form for the quantum Langevin equation,
\beq
\frac{dp}{dt} = -\frac{\partial{\cal H}}{\partial x} - F + {\mathfrak F},
\label{13a}
\eeq
\beq
\frac{dx}{dt} = \frac{\partial{\cal H}}{\partial p},
\label{13b}
\eeq
where $F$ is the dissipative force, that depends on $p$ and possibly on $x$,
${\mathfrak F}(t)$ is a fluctuating force
of the stochastic nature, and ${\cal H}$ is the Hamiltonian operator
\beq
{\cal H} = \frac{p^2}{2m}  + V(x).
\eeq
The quantities $p$ and $x$ appearing in equations (\ref{13a}) and (\ref{13b})
are quantum operators acting on a Hilbert space, and so are the dissipative
force $F$ and the Hamiltonian ${\cal H}$. However, the noise ${\mathfrak F}$
is a scalar and assumed to be complex.

In usual treatments of the Langevin equation,
the dissipative force $F$ is considered to be proportional to $p$,
but here we relax this condition and 
let $F$ be a generic function of $p$ and $x$, to be found,
which may not be Hermitian. If that is the case we see that 
the operators $x$ and $p$ in equations (\ref{13a}) and
(\ref{13b}) may not be Hermitian either. 
We also allow the noise to be a complex scalar, 
\beq
{\mathfrak F} = {\mathfrak F}_1 + i {\mathfrak F}_2,
\eeq
where ${\mathfrak F}_1$ and ${\mathfrak F}_2$
are real with the properties $\la {\mathfrak F}_n(t)\ra=0$ and
\beq
\la {\mathfrak F}_n(t){\mathfrak F}_{n'}(t')\ra
= \Gamma \delta_{nn'}\,\delta(t-t').
\label{3bb}
\eeq 

%------------------------------------------------
\subsection{Evolution of the density operator}

The equations of motion (\ref{13a}) and (\ref{13b}) are understood 
as equations belonging to the Heisenberg representation, which are
characterized by the time dependency of operators. An equivalent representation
which will be pursued next 
is the one in which states rather than operators change in time. 
However, the states in this representation will not be 
expressed by state vectors, describing pure states,
because the time evolution of $x$ and $p$, given by equations (\ref{13a})
and (\ref{13b}), is not unitary. Instead, the states will be
expressed by the density operator $\rho$, describing mixed states.

To find the equation that gives the time evolution of
$\rho$, we proceed in a way similar to the procedure employed in the
classical case. We start by writing the discretized version of
equations (\ref{13a}) and (\ref{13b}),
\beq
\Delta p = -\frac{\partial{\cal H}}{\partial x} \Delta t
- F\Delta t + \xi\sqrt{\Gamma\Delta t},
\label{31a}
\eeq
\beq
\Delta x = \frac{\partial{\cal H}}{\partial p}\Delta t,
\label{31b}
\eeq
where $\xi=\xi_1+i\xi_2$
and $\xi_1$ and $\xi_2$ are independent and uncorrelated random variables with
central Gaussian distributions with variance equal to 1/2. 

Next we consider the time evolution of an operator ${\cal F}(x,p)$.
The increment of ${\cal F}$ during an interval of time $\Delta t$ is
\[
\Delta {\cal F} = 
  \left(\frac{\partial{\cal F}}{\partial x},\Delta x\right) 
+ \left(\frac{\partial{\cal F}}{\partial x^\dagger},\Delta x^\dagger\right)
+ \left(\frac{\partial{\cal F}}{\partial p},\Delta p\right)
+ \left(\frac{\partial{\cal F}}{\partial p^\dagger},\Delta p^\dagger\right)
\]
\beq
+ \frac{\Gamma}2\left(
  \xi^2 \frac{\partial^2{\cal F}}{\partial p^2}
+2\xi^*\xi\frac{\partial^2{\cal F}}{\partial p\partial p^\dagger}
+ (\xi^*)^2 \frac{\partial^2{\cal F}}{\partial (p^\dagger)^2}\right)\Delta t.
\eeq
The notation $(,)$ used for a derivative is explained in the Appendix.

Replacing $\Delta x$, $\Delta p$,
given by the equations (\ref{31a}) and (\ref{31b}), and
after integrating in $\xi$ and dividing by $\Delta t$, we reach the
following equation by taking the limit $\Delta t\to0$,
\[
\frac{d\la{\cal F}\ra}{dt}
= \la\{{\cal F},{\cal H}\}\ra
- \la\left(\frac{\partial{\cal F}}{\partial p},F \right) \ra
\]
\beq
+ \la\{{\cal F},{\cal H}\}^\dagger\ra
- \la\left(\frac{\partial{\cal F}}{\partial p^\dagger},F^\dagger \right)\ra
+ \Gamma\la\frac{\partial^2{\cal F}}{\partial p\partial p^\dagger} \ra,
\label{47}
\eeq
where
\beq
\{{\cal F},{\cal H}\} = 
  \left(\frac{\partial{\cal F}}{\partial x}, \frac{\partial{\cal H}}{\partial p}\right)
- \left(\frac{\partial{\cal F}}{\partial p}, \frac{\partial{\cal H}}{\partial x}\right),
\eeq
and we are using the notation $\la{\cal A}\ra={\rm Tr}{\cal A}\rho$
for the average of ${\cal A}$ with respect to the density operator $\rho$.

Employing the results
\beq
{\rm Tr}\{{\cal F},{\cal H}\}\rho = - {\rm Tr}{\cal F}\{\rho,{\cal H}\},
\eeq
\beq
{\rm Tr}\left(\frac{\partial{\cal F}}{\partial p},F \right) \rho
= - {\rm Tr} {\cal F}\left(\frac{\partial\rho}{\partial p},F \right)
- {\rm Tr}{\cal F}\left(\rho\,\frac{\partial F}{\partial p}\right),
\eeq
where we have considered boundary conditions such that $\rho$
vanishes, and considering that
the function ${\cal F}$ is arbitrary, we may write the following
equation that gives the evolution of the density operator,
\[
\frac{\partial\rho}{\partial t} =
- \{\rho,{\cal H}\}
+ \left(\frac{\partial\rho}{\partial p},F\right)
+ \rho\frac{\partial F}{\partial p}
\]
\beq
- \{\rho,{\cal H}\}^\dagger
+ \left(\frac{\partial\rho}{\partial p^\dagger},F^\dagger\right)
+ \frac{\partial F^\dagger}{\partial p^\dagger}\rho
+ \Gamma \frac{\partial^2 \rho}{\partial p\partial p^\dagger}.
\eeq

%------------------------------------------------
\subsection{Thermodynamic equilibrium}

As we have seen above the operators $x$ and $p$, solutions
of equations (\ref{13a}) and (\ref{13b}), may not be Hermitian 
and in fact they remain non-Hermitian even in the stationary state.
In this sense they cannot be understood as the position
and momentum of a particle, which according to the principles
of quantum mechanics are observables and as a consequence should be Hermitian operators.
This nuisance is solved by requiring
that $p$ and $x$, the solution of equations (\ref{13a}) and (\ref{13b}), be 
normal operators, which are operators holding the following property
\beq
p^\dagger p=p p^\dagger, \qquad\qquad
x^\dagger x=x x^\dagger,
\eeq
which allows us to define the Hermitian operators
$\overline{x}$ and $\overline{p}$ by  
\beq
\overline{x} = (x^\dagger x)^{1/2},  \qquad\qquad
\overline{p} = (p^\dagger p)^{1/2}.
\label{19}
\eeq
We then {\it postulate} that the observables corresponding to the
position and momentum of the particle are $\x$ and $\p$, respectively,
and not properly $x$ and $p$.

In the stationary state we demand that the density operator is of the Gibbs type
\beq
\rho_0 = \frac1\zeta e^{-\beta{\cal H}}.
\label{43}
\eeq
In accordance with our postulate above, the Hamiltonian in 
equation (\ref{43}) will be written in terms of $\x$ and $\p$, that is,
\beq
{\cal H} = \frac{\p^2}{2m} + V(\x).
\label{44}
\eeq

To determine $F$, we proceed as follows. First we observe that
$\{\rho_0,{\cal H}\}=0$, a result that allows us
to reach the following equations that determine $F$,
\beq
\left(\frac{\partial\rho}{\partial p},F\right) + \rho\frac{\partial F}{\partial p}
+ \frac{\Gamma}2 \frac{\partial^2 \rho}{\partial p\partial p^\dagger} = 0.
\eeq
The solution of this equation is 
\beq
F = - \frac\Gamma2 \rho_0^{-1} \,\frac{\partial\rho_0}{\partial \p}
= - \frac\Gamma2 e^{\beta{\cal H}} \frac{\partial}{\partial \p} e^{-\beta{\cal H}},
\label{60}
\eeq
which is the desired dissipation-fluctuation relation.

The result (\ref{60}) can be expressed in the following form
\beq
F = \frac\Gamma2 \left(\beta \frac{\partial{\cal H}}{\partial \p}
- \frac{\beta^2}{2!} [\frac{\partial{\cal H}}{\partial \p},{\cal H}]
+ \frac{\beta^3}{3!} [[\frac{\partial{\cal H}}{\partial \p},{\cal H}],{\cal H}]
+ \ldots \right).
\eeq
Taking into account equation (\ref{44}), this expression reduces to
\beq
F = \frac{\Gamma \beta}{2m} \left(\p - \frac{\beta}{2!} [\p,{\cal H}]
+ \frac{\beta^2}{3!} [[\p,{\cal H}],{\cal H}]
+ \ldots \right),
\label{66}
\eeq
which is an equivalent form of writing the dissipation-fluctuation relation
(\ref{60}).

%--------------------------------------------------------------------
\section{Harmonic oscillator}

\subsection{Equilibrium}

Here we treat the harmonic oscillator for which
${\cal H}=p^2/2m+m\omega^2x^2/2$.
The dissipation force $F$ turns out to be linear in $p$ and $x$
which can be determined from (\ref{66}) as shown in the Appendix.
The result is
\beq
F = \gamma (ap - i b x), \qquad\qquad \gamma = \frac{\Gamma\beta}{2m}, 
\eeq
where
\beq
a = \frac1{\beta\hbar\omega}\sinh\beta\hbar\omega,
\qquad\qquad  b = \frac{m}{\beta \hbar}(\cosh\beta\hbar\omega - 1). 
\label{64} 
\eeq
We use the Langevin equations (\ref{13a}) and (\ref{13b}),
which for the harmonic oscillator reads
\beq
\frac{dp}{dt} = -m\omega^2 x - \gamma(ap-ibx) + {\mathfrak F},
\label{58a}
\eeq
\beq
\frac{dx}{dt} = \frac{p}{m}.
\label{58b}
\eeq

From these equations we obtain
\beq
\frac{d}{dt} \la p^\dagger p\ra = -m\omega^2 \la x^\dagger p + p^\dagger x \ra
- 2 \gamma a \la p^\dagger p\ra - i\gamma b \la x^\dagger p - p^\dagger x\ra
+ \frac{2m\gamma}\beta,
\label{36a}
\eeq 
\beq
\frac{d}{dt} \la x^\dagger x\ra = \frac1m \la x^\dagger p + p^\dagger x\ra,
\label{36b}
\eeq
\beq
\frac{d}{dt} \la x^\dagger p\ra
= \frac1m \la p^\dagger p\ra - m\omega^2 \la x^\dagger x\ra
- \gamma a \la x^\dagger p\ra + i\gamma b \la x^\dagger x\ra,
\label{36c}
\eeq
\beq
\frac{d}{dt} \la p^\dagger x \ra =
\frac1m \la p^\dagger p\ra - m\omega^2 \la x^\dagger x\ra
- \gamma a \la p^\dagger x\ra -  i\gamma b \la x^\dagger x\ra.
\label{36d}
\eeq
Summing and subtracting the last two equations give
\beq
\frac{d}{dt} \la x^\dagger p + p^\dagger x \ra = \frac2m \la p^\dagger p\ra
- 2 m\omega^2 \la x^\dagger x\ra- \gamma a \la x^\dagger p + p^\dagger x\ra,
\label{36e}
\eeq
\beq
\frac{d}{dt} \la x^\dagger p - p^\dagger x \ra =
- \gamma a \la x^\dagger p - p^\dagger x\ra + 2 i\gamma b \la x^\dagger x\ra.
\label{36f}
\eeq

The equations (\ref{36a}), (\ref{36b}), (\ref{36e}), and (\ref{36f}) 
form a set of linear closed equations for the four correlations
$\la p^\dagger p\ra$, $\la x^\dagger x\ra$, $\la x^\dagger p +  p^\dagger x\ra$
and $\la x^\dagger p -  p^\dagger x\ra$, and can be solved as functions
of time. Here however we will focus on the steady state solution,
which is given by the equations
\beq
\la x^\dagger p + p^\dagger x\ra = 0,
\eeq
\beq
\frac1m \la p^\dagger p\ra = m\omega^2 \la x^\dagger x\ra 
= \frac{a}{\beta(a^2 - b^2/m^2\omega^2)},
\eeq
\beq
\la x^\dagger p - p^\dagger x\ra
= \frac{2i b}{\beta m\omega^2 (a^2-b^2/m^2\omega^2)}.
\eeq
Using the values of $a$ and $b$ given by (\ref{64}), we reach the results
\beq
\frac1m \la p^\dagger p\ra = m\omega^2 \la x^\dagger x\ra 
= \hbar\omega\left(\frac{1}{e^{\beta\hbar\omega}-1} + \frac12\right),
\label{70}
\eeq
\beq
\la x^\dagger p - p^\dagger x\ra = i\hbar,
\label{71}
\eeq
which are the desired results, if we bear in mind that
$p^\dagger p = \p^2$ and $x^\dagger x = \x^2$.

%------------------------------------------------
\subsection{Time dependent solution}

The set of equations (\ref{58a}) and (\ref{58b})
are a set of linear differential equations and can thus be
solved for any given time dependent function ${\mathfrak F}$.
From the solution for $x(t)$ and $p(t)$ the
covariances are readily obtained. Here, we take a distinct approach
and instead we solve the equations
for the covariances themselves, which are given by
equations (\ref{36a}), (\ref{36b}), (\ref{36c}), and (\ref{36d}).
To this end we first observe that these equations can be written
in terms of a matrix differential equation.
Defining $X=\la x^\dagger x\ra$, $Y=\la p^\dagger p\ra$,  
$Z=\la x^\dagger p\ra$, then
\beq
\frac{dY}{dt}  = -m\omega^2 (Z+Z^*) - 2 \gamma a Y - i\gamma b (Z-Z^*)
+ \frac{2m\gamma}{\beta},
\label{37a}
\eeq 
\beq
\frac{dX}{dt} = \frac1m(Z+Z^*),
\label{37b}
\eeq
\beq
\frac{dZ}{dt} = \frac1m Y - m\omega^2 X - \gamma a Z + i\gamma b X,
\label{37c}
\eeq
\beq
\frac{dZ^*}{dt} = \frac1m Y - m\omega^2 X - \gamma a Z^* -  i\gamma b X.
\label{37d}
\eeq
The time evolution of the covariance matrix
\beq
Q = \left(
\begin{array}{cc}
X     & Z \\
Z^*   & Y \\
\end{array}
\right)
\eeq
is easily obtained from these equations and is given by
\beq
\frac{dQ}{dt} = Q M + M^\dagger Q + \Omega,
\label{68}
\eeq
where the matrices $M$ are $\Omega$ are
\beq
M = \left(
\begin{array}{cc}
0      & -m\omega^2 +i\gamma b \\
1/m    & -\gamma a \\
\end{array}
\right), \qquad\qquad \Omega = \left(
\begin{array}{cc}
0    & 0 \\
0    & 2m\gamma/\beta\\
\end{array}
\right),
\eeq
and $M^\dagger$ is the transpose conjugate of $M$.

Next we set up a matrix $B$ whose columns are the right eingevectors
of the matrix $M$. This matrix and its inverse are
\beq
B = \left(
\begin{array}{cc}
-\lambda_2 &  \lambda_1 \\
  1/m      & -1/m      \\
\end{array}
\right),
\qquad\qquad
B^{-1} = \frac{m}{\lambda_1-\lambda_2} \left(
\begin{array}{cc}
 1/m   &  \lambda_1\\
 1/m   &  \lambda_2 \\
\end{array}
\right),
\eeq
where $\lambda_1$ and $\lambda_2$ are the eigenvalues of
$M$ and the roots of
\beq
m \lambda^2 + \gamma m a \lambda + (m\omega^2 - i\gamma b) = 0.
\eeq
The matrix $B$ together with its inverse $B^{-1}$
diagonalizes $M$, that is, $B^{-1} M B=D$, where $D$ is diagonal with elements
$\lambda_1$ and $\lambda_2$. 

As to the matrix $M^{\dagger}$, it is not diagonalized by the matrix $B$
but it is diagonalized by the transpose conjugate of $B$. Indeed,
taking the transpose conjugate of $B^{-1} M B=D$, we find the result
$A M^\dagger A^{-1} = D^\dagger$, where $A=B^\dagger$, and $D^\dagger$
is the diagonal matrix with elements $\lambda_1^*$ and $\lambda_2^*$.
The matrix $A$ and its inverse are
\beq
A = \left(
\begin{array}{cc}
-\lambda_2^* &  1/m \\
 \lambda_1^*      & -1/m      \\
\end{array}
\right),
\qquad\qquad
A^{-1} = \frac{m}{\lambda_1^*-\lambda_2^*} \left(
\begin{array}{cc}
 1/m   &  1/m \\
 \lambda_1^*   &  \lambda_2^* \\
\end{array}
\right).
\eeq

Multiplying the equation (\ref{68}) at the left by $A$ and
at the right by $B$, we reach the equation
\beq
\frac{dQ'}{dt} = Q' D + D^\dagger  Q'+ \Omega',
\label{68a}
\eeq
where $Q'=AQB$ and $\Omega'=A\Omega B$.
Denoting by $X'$, $Y'$, $Z'$, and $(Z')^*$ the elements of
the matrix $Q'$, the equation (\ref{68a}) is written in
a explicit form as
\beq
\frac{dX'}{dt} = (\lambda_1+\lambda_1^*) X' + \frac{2\gamma}{m\beta},
\eeq
\beq
\frac{dY'}{dt} = (\lambda_2+\lambda_2^*) Y' + \frac{2\gamma}{m\beta},
\eeq
\beq
\frac{dZ'}{dt} = (\lambda_1^*+\lambda_2) Z' - \frac{2\gamma}{m\beta},
\eeq
\beq
\frac{d(Z')^*}{dt} = (\lambda_1+\lambda_2^*) (Z')^* - \frac{2\gamma}{m\beta}. 
\eeq
The solutions of these equations are readily obtained,
\beq 
X' =  X_0 e^{(\lambda_1+\lambda_1^*)t} - \frac{2\gamma}{m\beta(\lambda_1+\lambda_1^*)},
\eeq
\beq 
Y' =  Y_0 e^{(\lambda_2+\lambda_2^*)t} - \frac{2\gamma}{m\beta(\lambda_2+\lambda_2^*)},
\eeq
\beq 
Z' =  Z_0 e^{(\lambda_1^*+\lambda_2)t} + \frac{2\gamma}{m\beta(\lambda_1^*+\lambda_2)},
\eeq
\beq 
(Z')^* =  Z_0^* e^{(\lambda_1+\lambda_2^*)t}
+ \frac{2\gamma}{m\beta(\lambda_1+\lambda_2^*)}.
\eeq

The relation $Q'=AQB$ gives $Q=A^{-1}Q'B^{-1}$ from which follows
\beq
X = \frac{1}{|\lambda_1-\lambda_2|^2}(X' + Y' + Z' + (Z')^*),
\label{72a}
\eeq
\beq
Y = \frac{m^2}{|\lambda_1-\lambda_2|^2}
(|\lambda_1|^2 X' + |\lambda_2|^2 Y'
+\lambda_1^* \lambda_2 Z' + \lambda_2^*\lambda_1 (Z')^*),
\label{72b}
\eeq
\beq
Z = \frac{m}{|\lambda_1-\lambda_2|^2}
(\lambda_1 X' + \lambda_2 Y' + \lambda_1 (Z')^* + \lambda_2 Z'),
\label{72c}
\eeq
\beq
Z^* = \frac{m}{|\lambda_1-\lambda_2|^2}
(\lambda_1^* X' + \lambda_2^* Y' + \lambda_1^* Z' + \lambda_2^* (Z')^*),
\label{72d}
\eeq
which are the relations between the covariances $X$, $Y$, $Z$, and $Z^*$,
and $X'$, $Y'$, $Z'$, and $(Z')^*$. The replacement of 
the later variables into the equations (\ref{72a}), (\ref{72b}), 
(\ref{72c}), and (\ref{72d}), gives the covariances as functions of time.   

When $t\to\infty$ its possible to show that these covariances reaches
the equilibrium values given by equations (\ref{70}) and (\ref{71}).
These values are approached exponentially with a correlation time
which is the reciprocal of the real part of the eigenvalues.

%--------------------------------------------------------------------
\section{Conclusion}

We have proposed a quantum Langevin equation that is very distinct
from previous approaches. The distinction concerns both the dissipation
and fluctuation forces. The fluctuation force is similar to that employed
in the classical case. It is a white noise with a covariance proportional
to the temperature and is the same for any system. The fluctuation force
is a scalar quantity, instead of an operator, which we found it convenient
to consider as a complex quantity. 
The dissipation force is not restrict to
be proportional to the velocity as in the usual approaches to the Langevin equation.
In our approach it is determined as to guarantee that the stationary
state is given by a density operator of the Gibbs canonical type.
The system approaches equilibrium or in other words the system thermalizes.
This result is general for any systems as long as one uses the
dissipation force given by relation (\ref{66}).
To find the relation between dissipation and fluctuation that leaded
to relation (\ref{66}), we have derived an equation that gives the
time evolution of the density operator, which turns out to be a
quantum Fokker-Planck-Kramers equation, proposed previously.

The Langevin equation is written in terms of the operators $x$ and $p$
which are not, in general, hermitian operator. To overcome this 
problem we have postulated that the actual position and momentum are
the the operators $\x=(x^\dagger x)^{1/2}$ and $\p=(p^\dagger p)^{1/2}$,
respectively.
In equilibrium the density operator is of the Gibbs canonical type
with a Hamiltonian written in terms of $\x$ and $\p$. 

The position and momentum operator obey, in equilibrium, the commutation
relation in the form $x^\dagger p - p^\dagger x=i\hbar$, although 
this relation is not obeyed out of equilibrium. Although the
fulfillment of the commutation relation, at any time,
is a basic assumption in
other approaches to the Langevin equation, our understanding is that
this demand is only necessary when the dynamics is unitary,
which is not the case of the Langevin dynamics. This situation is analogous
to the Hamiltonian dynamics in the classical case for which
the Poisson brackets of the position and momentum in relation
to the initial position and momentum, at any  
time, is always equal to one. In the Langevin dynamics, this may not be
required.

Finally, we remark that the quantum Fokker-Planck-Kramers equation
derived here from the Langevin equation is very similar to that
obtained by the canonical quantization of the classical 
Fokker-Planck-Kramers equation \cite{oliveira2016}, which has
been applied to the calculation of the transport properties
of a chain of coupled harmonic oscillators and of a bosonic
chain \cite{oliveira2017,oliveira2018}.

%--------------------------------------------------------------------
\appendix

%--------------------------------------------------------------------
\section{}

Let $X$ be a matrix that depends on a parameter $s$. The derivative of
$X$ with respect to $s$ is defined by
\beq
\frac{\partial X}{\partial s} = \lim_{\varepsilon\to0}\frac{X(s+\varepsilon)-X(s)}{\varepsilon}.
\label{61}
\eeq
If a matrix $A$ is a function of another matrix $X$, then the derivative
of $A$ with respect to $X$ is the matrix defined by
\beq
\frac{\partial A}{\partial X} = \lim_{\varepsilon\to0}\frac{A(X+\varepsilon I)-A(X)}{\varepsilon},
\label{63}
\eeq
where $I$ is the identity matrix.

If $I$ is replaced by a matrix $B$, the resulting derivative will depend on
$B$. In this case we use a distinct notation,
\beq
\left(\frac{\partial A}{\partial X},Y\right) = 
\lim_{\varepsilon\to0}\frac{A(X+\varepsilon Y)-A(X)}{\varepsilon}.
\label{65}
\eeq
If a series expansion of $A(X+Y)$ in powers of $Y$ is carried out,
the derivatives to be used are derivatives of this type.
Up to second order in $Y$,
\beq
A(X+B) = A(X) + \left(\frac{\partial A}{\partial X},Y\right) 
+ \frac12 \left(\frac{\partial}{\partial X}
\left(\frac{\partial A}{\partial X},Y\right),Y\right).
\eeq
Notice that this derivative is not in general equal to $(\partial A/\partial X)$
multiplied by $Y$. 
The derivative given by (\ref{63}) and that given by (\ref{65}) are related by
\beq
\frac{\partial A}{\partial X} = \left(\frac{\partial A}{\partial X},I\right).
\eeq
As a simple example of the derivatives defined above, let us consider $A=X^2$.
Then the usual derivative given by (\ref{63}) is $2 X$ whereas that given by
(\ref{65}) is $YX+XY$.

If $A(Y)$ depends on $Y(X)$ that depends on $X$, the derivative of $A$
with respect to $X$ is given by the chain rule is
\beq
\frac{\partial A}{\partial X}
= \left(\frac{\partial A}{\partial Y},\frac{\partial Y}{\partial X}\right).
\eeq

%--------------------------------------------------------------------
\section{}

We calculate here the expression
\beq
g = p - \frac{\beta}{2!} [p,{\cal H}]
+ \frac{\beta^2}{3!} [[p,{\cal H}],{\cal H}] + \ldots,
\label{66a}
\eeq
for the harmonic oscillator, for which
\beq
{\cal H} = \frac{p^2}{2m} + \frac{m\omega^2}2 x^2.
\eeq
To this end we define the quantities $A_n$ by
\beq
A_{n+1} = [A_n,{\cal H}] \qquad\qquad A_0=p,
\eeq
so that 
\beq
g = \sum_{n=0}^\infty \frac{(-\beta)^n}{(n+1)!}A_n.
\eeq

Now, using the commutation relation $[x,p]=i\hbar$, we find
\beq
[p,{\cal H}] = - i \hbar m\omega^2 x,
\eeq
and
\beq
[x,{\cal H}] =  i  \hbar \frac{p}m,
\eeq
which allows us to conclude draw the following conclusion:
if $n$ is even then  $A_n=a_n p$, and if $n$ is odd then $A_n=b_n x$ 
Replacing these two results in the recursive definition of $A_n$,
we find $b_{n+1} = - i \hbar m\omega^2 a_n$ for $n$ even, 
and $a_{n+1} = (i\hbar/m) b_n$ for n odd. These results
give $a_{n+2} = (\hbar\omega)^2 a_n$ and $b_{n+2} = (\hbar\omega)^2 b_n$
from which follows that
\beq
a_n = (\hbar\omega)^n a_0,
\eeq
\beq
b_{n+1} = (\hbar\omega)^n b_1, 
\eeq
and expression (\ref{66a}) becomes
\beq
g = \sum_{n\,{\rm even}} \frac{(\beta\hbar\omega)^n}{(n+1)!} a_0 p
- \sum_{n\,{\rm odd}} \frac{\beta^n(\hbar\omega)^{n-1}}{(n+1)!} b_1 x,
\eeq
or
\beq
g = \frac{\sinh\beta\hbar\omega}{\beta\hbar\omega}  a_0 p
- \frac{1}{\beta(\hbar\omega)^2}(\cosh\beta\hbar\omega -1) b_1 x.
\eeq
Considering that $a_0=1$ and $b_1=-i\hbar m\omega^2$,
we reach the result
\beq
g = \frac{\sinh\beta\hbar\omega}{\beta\hbar\omega} p
+ \frac{i m}{\beta\hbar}(\cosh\beta\hbar\omega -1) x,
\eeq
which we write as
\beq
g = a p + i b x,
\eeq
where
\beq
a = \frac{\sinh\beta\hbar\omega}{\beta\hbar\omega},
\qquad\qquad
b = \frac{m}{\beta\hbar}(\cosh\beta\hbar\omega -1).
\eeq

%--------------------------------------------------------------------


\section*{References}

\begin{thebibliography}{99}

\bibitem{kampen1981} N. G. van Kampen, {\it Sochastic Processes in
Physics and Chemistry}, (North-Holland, Amsterdam, 1981).

\bibitem{risken1989} H. Risken, {\it The Fokker-Planck Equation},
(Springer, Berlin, 1989); 2nd. ed. 

\bibitem{gardiner2009} C. W. Gardiner, {\it Stochastic Methods},
(Springer, Berlin, 2009); 4th. ed.

\bibitem{coffey1996} W. T. Coffey, Yu. P. Kalmykov, and J. T. Waldron,
{\it The Langevin Equation}, (World Scientific, Singapore, 1996).

\bibitem{carmichael1999} H. J. Carmichael, {\it Statistical Methods
in Quantum Optics 1}, (Springr, Heidelberg, 1999)

\bibitem{breuer2002}
H.-P. Breuer and F. Petruccione,  {\it The Theory of Open Quantum Systems},
(Oxford University Press, Oxford, 2002).

\bibitem{gardiner2010} C. W. Gardiner and P. Zoller, {\it Quantum Noise}
(Springer, Berlin, 2010).

\bibitem{weiss2012} U. Weiss, {\it Quantum Dissipative Systems},
(World Scientific, Singapore, 2012); 4th ed. 

\bibitem{caldeira2014} A. O. Caldeira, {\it An Introduction to Macroscopic Quantum
Phenomena and Quantum Dissipation}, (Cambridge University Press, Cambridge, 2014).

\bibitem{araujo2019} R. Ara\'ujo, S. Wald, and M. Henkel,
''Axiomatic construction of quantum Langevin equations'',
J. Stat. Mech. (2019) 053101.

\bibitem{senitzky1960} I. R. Senitzky,
''Dissipation in quantum mechanics. The Harmonic oscillator. II'',
Phys. Rev. {\bf 119}, 670 (1960).

\bibitem{senitzky1961} I. R. Senitzky,
''Dissipation in quantum mechanics. The Harmonic oscillator'',
Phys. Rev. {\bf 124}, 642 (1961).

\bibitem{frenkel2012} J. Frenkel and J. C. Taylor, 
''Condition for the validity of the quantum Langevin equation'',
Phys. Rev. E {\bf 85}, 011135 (2012).

\bibitem{ford1965} G. W. Ford, M. Kac, and P. Mazur,
''Statistical mechanics of assemblies of coupled oscillators'',
J. Math. Phys. {\bf 6}, 504 (1965).

\bibitem{pasquale1984} F. de Pasquale, P. Ruggiero, and M. Zannetti,
''Noise spectrum in the quantum Langevin equation'',
J. Phys. A: Math. Gen. {\bf 17}, 1489 (1984). 

\bibitem{ford1987} G. W. Ford and M. Kac,  
''On the quantum Langevin equation'',
J. Stat. Phys. {\bf 46}, 803 (1987).

\bibitem{ford1988} G. W. Ford, J. T. Lewis, and R. F. O’Connell,
''Quantum Langevin equation'',
Phys. Rev. A {\bf 37}, 4419 (1988).

\bibitem{zurcher1990} U. Z\"urcher and P. Talkner,
''Quantum-mechanical harmonic chain attached to heat baths'',
Phys. Rev. A {\bf 42}, 3267, 3278 (1990).

\bibitem{benguria1981} R. Benguria and M. Kac,
''Quantum Lanagevin equation'',
Phys. Rev. Lett. {\bf 46}, 1 (1981).

\bibitem{maassen1984} H. Maassen,
''Return to thermal equilibrium by the solution of a quantum Langevin equation'',
J. Stat. Phys. {\bf 34}, 239 (1984).

\bibitem{bedeaux2001} D. Bedeaux and P. Mazur,
''Mesoscopic non-equilibrium thermodynamics for quantum systems'',
Physica A {\bf 298}, 81 (2001).

\bibitem{oliveira2016} M. J. de Oliveira,
''Quantum Fokker-Planck-Kramers equation and entropy production'',
Phys. Rev. E {\bf 94}, 012128 (2016).

\bibitem{oliveira2017} M. J. de Oliveira,
''Heat transport along a chain of coupled quantum harmonic oscillators'',
Phys. Rev. E {\bf 95}, 042113 (2017).

\bibitem{oliveira2018} M. J. de Oliveira,
''Stochastic quantum thermodynamics, entropy production, and
transport properties of a bosonic system'',
Phys. Rev. E {\bf 97}, 012105 (2018).


\end{thebibliography}
\end{document}